**ORIGINAL ARTICLE**

# Volume reduction of water samples to increase sensitivity for radioassay of lead contamination

A. Aguilar-Arevalo[1] · C. Canet[2] · M. A. Cruz-Pérez[2] · A. Deisting[3] · A. Dias[3] · J. C. D'Olivo[1] · F. Favela-Pérez[1,4] · E. A. Garcés[5] · A. González Muñoz[5] · J. O. Guerra-Pulido[1] · J. Mancera-Alejandrez[6] · D. J. Marín-Lámbarri[5] · M. Martinez Montero[1] · J. R. Monroe[3] · S. Paling[7] · S. J. M. Peeters[8] · P. R. Scovell[7] · C. Türkoğlu[8,9] · E. Vázquez-Jáuregui[5] · J. Walding[3]

**Abstract**
The World Health Organisation (WHO) presents an upper limit for lead in drinking water of 10 parts per billion ppb. Typically, to reach this level of sensitivity, expensive metrology is required. To increase the sensitivity range of low-cost devices, this paper explores the prospects of using a volume reduction technique of a boiled water sample doped with Lead-210 ($^{210}$Pb), as a means to increase the solute's concentration. $^{210}$Pb is a radioactive lead isotope and its concentration in a water sample can be measured with e.g. High Purity Germanium (HPGe) detectors at the Boulby Underground Germanium Suite. Concentrations close to the WHO limit have not been examined. This paper presents a measurement of the volume reduction technique retaining $99 \pm (9)\%$ of $^{210}$Pb starting from a concentration of $1.9 \times 10^{-6}$ ppb before reduction and resulting in $2.63 \times 10^{-4}$ ppb after reduction. This work also applies the volume reduction technique to London tap water and reports the radioassay results from gamma counting in HPGe detectors. Among other radio-isotopes, $^{40}$K, $^{210}$Pb, $^{131}$I and $^{177}$Lu were identified at measured concentrations of $2.83 \times 10^{3}$ ppb, $2.55 \times 10^{-7}$ ppb, $5.06 \times 10^{-10}$ ppb and $5.84 \times 10^{-10}$ ppb in the London tap water sample. This technique retained $90 \pm 50\%$ of $^{40}$K. Stable lead was inferred from the same water sample at a measured concentration of 0.012 ppb, prior to reduction.

✉ A. Dias
adriana.dias.2011@live.rhul.ac.uk

[1] Instituto de Ciencias Nucleares, Universidad Nacional Autónoma de México, Ciudad de México, CDMX, México

[2] Centro de Ciencias de la Atmósfera, Universidad Nacional Autónoma de México, Ciudad de México, CDMX 04510, México

[3] Royal Holloway, University of London, Egham Hill, UK

[4] Centro Atómico Bariloche, CNEA/CONICET/IB, Bariloche, Argentina

[5] Instituto de Física, Universidad Nacional Autónoma de México, A. P. 20-364, Ciudad de México, DF 01000, México

[6] Facultad de Ingeniería, Universidad Nacional Autónoma de México, Ciudad de México, México

[7] Boulby Underground Laboratory, Boulby Mine, Saltburn-by-the-Sea, UK

[8] Department of Physics and Astronomy, University of Sussex, Brighton, UK

[9] Present Address: AstroCeNT–Particle Astrophysics Science and Technology Centre, Warsaw, Poland

## Introduction

Lead (Pb) pollution is among the chief causes of years of productive life loss in the world, and it has been estimated that 1 in 3 children in the world has lead blood levels above the World Health Organisation (WHO) limit (UNICEF and Pure Earth 2020). A leading source of lead poisoning of children in low- and middle-income countries (LMICs) is pollution associated with improper recycling of lead-acid batteries. The cost of childhood lead exposure due to lost economic potential of these children over their lifetime is estimated to approach USD $1 trillion in LMICs (UNICEF and Pure Earth 2020).

The WHO limit for Pb in drinking water is 10 parts per billion ppb (Geneva: World Health Organization 2017). Reaching this level of sensitivity in water quality measurements typically requires dedicated and relatively expensive metrology (Water 2021). With the aim of developing low-cost, widely available metrology for lead in drinking water, we recently explored the potential of leveraging the precision calorimetry techniques of silicon (Si) detectors. This work demonstrated the capability of identifying Lead-210 $^{210}$Pb at the 100 ppb level in a sample (Aguilar-Arevalo et al. 2020b), through calorimetry and identification of characteristic gamma energies in a Si detector. While this technique is promising for dosimetry applications, it is not able to reach the WHO level in sensitivity. One way to increase the sensitivity range of any device is to increase the lead concentration in the sample. In this paper we explore heavy metal concentration increase by boiling doped water samples.

Volume concentration, by vaporization of acrylic, is a strategy developed to increase sensitivity to ultra-low concentrations of radio-isotopes by astroparticle physics experiments, and has been applied to measure the $^{210}$Pb content of acrylic (Nantais et al. 2013). Astroparticle physics experiments have used volume reduction techniques on solid samples. For the construction of the Sudbury Neutrino Observatory (SNO) experiment it was of critical importance to measure the detector acrylic's Uranium and Thorium chain radionuclide content in order to characterise and minimise the detector backgrounds (Boger et al. 2000). To achieve this, large samples of acrylic needed to be measured within a Broad Energy Germanium (BEGe) detector. To improve the efficiency of the assay campaign the acrylic was vapourised, reducing the sample to a char with a mass down to 15% (Holland and Hay 2001) of the original sample mass. Heavy metals associated with the $^{238}$U and $^{232}$Th decay chains remain in the char. The efficiency for this process to retain U and Th isotopes in the acrylic vaporization system of the SNO experiment was found to be >93% (Earle and Bonvin 1992). A vaporization approach using inductively coupled plasma mass spectrometry (ICP-MS) for the Jiangmen Underground Neutrino Observatory (JUNO) experiment achieved 75% efficiency for U and Th (cao et al. 2020). The SNO vaporization system was adapted and applied to measurement of $^{210}$Pb for the Dark matter Experiment using Argon Pulse-shape discrimination (DEAP-3600) experiment, where the efficiency for retention of lead in the char was measured in samples with independently-assayed $^{210}$Pb content to be 90–95% (Nantais 2014).

A similar technique has been employed when using $^{210}$Pb as tracer for lead in plants (Yanga and Appleby 2016). In Melucci et al. (2013), measurements of heavy metals were made, in matrices involved in food chain. Tea leaves were analysed with voltammetric procedures. Lead nitrate, Pb(II), was found at concentrations between 0.19 and 1.91 μg/g, which is comparable with other studies of the same element. This method was validated with electrothermal atomic absorption spectroscopy and showed good agreement between the different techniques. Nevertheless, it also does not reach the WHO level in sensitivity.

Here we apply the principle of reducing measurement samples in volume in order to increase the concentration of heavy metals for measurement. However, instead of vapourising solid samples we boil water samples to reduce their volume by evaporation. Our aim was to develop a method suitable for use in the field, without specialized equipment. For this experiment we have used water, spiked with a known amount of $^{210}$Pb. The molecular weight of $^{210}$Pb is similar to lead and the isotope can thus serve as a test bench for lead retention. Afterwards, the volume-reduced water samples are assayed with HPGe detectors to determine their radioisotope contents in order to asses the level at which these isotopes are retained. This paper reports the procedure for volume reduction in Sect. 2.1 and the efficiency for retaining lead in Sect. 3.1.1. This work demonstrates that the volume reduction technique retains >99% of $^{210}$Pb whilst increasing the $^{210}$Pb concentration to a level measurable via calorimetry in Si detectors (Aguilar-Arevalo et al. 2020b).

In a second experiment, using London tap-water, the method was employed. Tap water samples are reduced and assayed with a BEGe detector in order to determine the radioisotope content of the tap-water and to assess the isotopic retention efficiency, reported in Sect. 3.2.

## Methodology

Two different methods were employed to reduce the volume of water samples. The first method uses an enclosed system, which allows for full control of the exposure of the sample to the atmosphere throughout the volume reduction process. This is suitable for a laboratory environment. The second uses an open system, in which the evaporation escapes into the atmosphere, more similarly to how such a volume reduction method could be employed in the field without specialist equipment.

Using volume-reduction as a means to enhance radio-assay sensitivity requires knowledge of the retention of the isotopes of interest throughout the volume-reduction procedure. This was measured for both enclosed and open volume reduction procedures. The enclosed volume reduction used a source of known $^{210}$Pb activity. The open volume reduction procedure uses naturally occurring $^{40}$K to assess retention efficiency.

Both enclosed and open methods were studied inside an ESCO Laboratory fume hood, with an airflow speed between 1.2 and 1.5 ms$^{-1}$, in order to control the exhaust path of $^{210}$Pb-spiked samples.

### Enclosed system procedure

A $^{210}$Pb calibration source was prepared as follows. A standard liquid source with activity 518 ± 58 Bq dissolved in 1

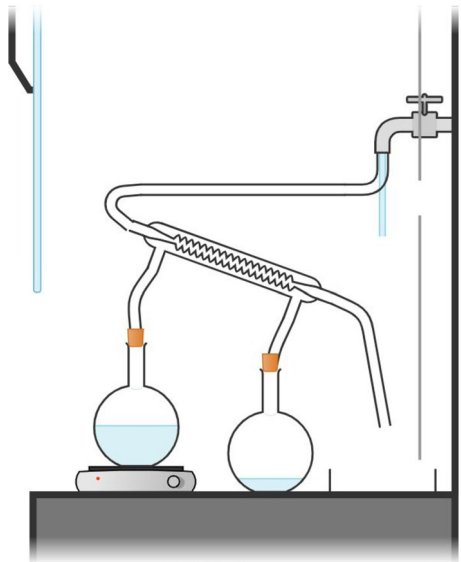

**Fig. 1** Schematic of enclosed system set up, detailing connections between the condenser and the two flasks

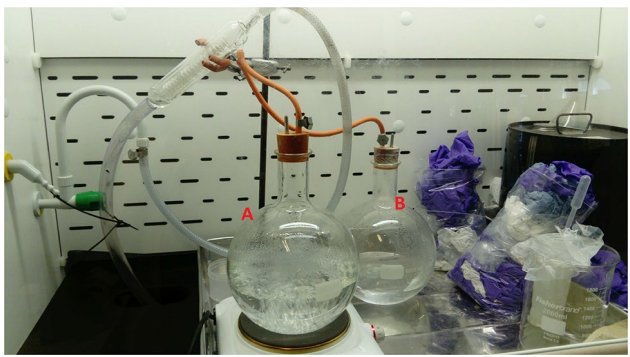

**Fig. 2** Photograph of the enclosed system during a boil, showing the flask undergoing volume-reduction, flask A, and the flask with the condensed liquid, flask B

mol of nitric acid, was purchased from Eckert and Ziegler Isotope Products (Eckert and Ziegler 2021). The mass of this standard source was $5.121 \pm 0.002$ g. A sample of $1.66 \pm 0.01$ g was removed from the standard source vial, and diluted in $4236 \pm 1$ g of ultra pure water water (UPW). This produced a diluted $^{210}$Pb calibration source for the measurements reported here, with activity of $167.9 \pm 18.8$ Bq (calculated as $518 \text{ Bq} \times \frac{1.66 \text{ g}}{5.121 \text{ g}}$).

To produce volume-reduced samples, a UPW sample, spiked with the calibration source, is boiled in an enclosed system. A schematic and photograph of the enclosed system are shown in Figs. 1 and 2, respectively.

In the enclosed system volume reduction apparatus, two DURAN 5000 ml round bottom flasks were connected on both sides of a coil condenser. The flask containing the UPW sample to be volume-reduced (flask A) was placed on top of a small electric hob and was connected to the inlet of the condenser. The outlet of the condenser was connected to the inlet of the flask for collection of the condensed solution (flask B). These connections were made using rubber tubing pipes with 6.74 mm inner diameter and the flasks were sealed with rubber bung stoppers to ensure no evaporated or liquid solution would leave the enclosed system. UPW to be used in the reduction was stored in 1000 ml Pyrex® flasks. Prior to reduction, the flasks and the condenser were cleaned with decon® 90 and ethanol. Throughout the reduction, samples were taken from flask A and flask B. These samples were contained in identical pots made of Polypropylene and with dimensions of 8.9 cm outer diameter and 6.4 cm height. They were also cleaned, prior to reduction, with UPW and isopropanol.

To begin the reduction, the cold water tap was turned on and adjusted until flow was steady and did not overflow the sink. The electric hob was turned on and the set up was monitored until it started boiling vigorously. When this point was reached, the hob was turned down until the UPW was observed to be simmering. The system was monitored every 30 min. When the volume of UPW in flask A was less than 1l, the hob and the cold water tap were turned off and the flasks were left to cool. Once cool, the bung from flask A was removed and more UPW was added. In the instance when the volume of flask B reached 5l, the hob and the cold water tap were also turned off, and the flasks were allowed to cool off. Following this, the bung was removed and the contents of flask B were deposited into a 20l polyethylene container of condensed UPW. This procedure was repeated until all the desired samples had been obtained.

With enclosed system reduction, samples with volume reduction ratios of 1:1, 1:10, 1:25 and 1:100 were prepared. The enclosed system reduction samples are listed in Table 1. The first sample was prepared from the input UPW (Blue I). Then, the non-reduced UPW was spiked with 1.66 g $^{210}$Pb solution, and a sample was extracted (Red II). The first reduction 1:10 (Green III) followed. After this reduction, Black IV was taken from the condensate flask. A sample was taken from Flask A after the second reduction reached 1:25 (Green V), and a final sample when the third reduction reached 1:100 (Blue VI). After this final reduction, Sample X was taken from the condensate flask. Finally, the remaining mass of the Eckert and Ziegler source solution, 3.33 g, was diluted into the remaining UPW, from which Sample IX was taken.

## Open system

For the open system reduction, 5 tap water samples of the following volume reduction ratios were prepared: 1:1, 1:10,

**Table 1** Corresponding sample and original mass values for different reductions

| Sample name | Description | Reduction | Sample mass | Original mass |
| --- | --- | --- | --- | --- |
| Blue I | UPW | – | 201 | n/a |
| Red II | UPW + $^{210}$Pb | 1:1 | 200 | 4236 |
| Green III | 1st reduction | 1:10 | 20 | 700 |
| Black IV | Condensate 1 | – | 200 | n/a |
| Green V | 2nd reduction | 1:25 | 20 | 233 |
| Blue VI | 3rd reduction | 1:100 | 191 | 26193 |
| Sample IX | Final Source sample | – | 200 | n/a |
| Sample X | Condensate 2 | – | 205 | n/a |

Samples related to reductions were prepared from the boiling flask. Samples referred to as "Condensate" in Description column were prepared from the condensate flask. Sample Mass values are contained in the Original Mass values. Both are quoted in (g). Sample and Original masses were measured with a set of scales with an uncertainty of ±1 g 1:100, 1:1000 and 1:5000. A sample of limescale extracted from the condensate was also prepared. To achieve this, two Pyrex® trays were positioned inside the fume hood, on top of two electric hobs and tap water was deposited inside the trays (Figs. 3 and 4). The trays had dimensions of 0.27 m × 0.4 m × 0.055 m ($L \times W \times H$).

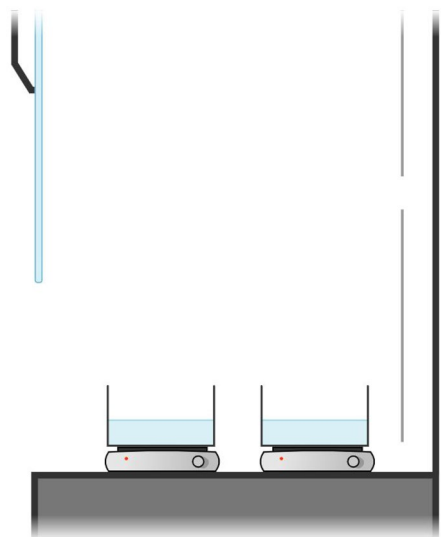

**Fig. 3** Schematic of open system set up

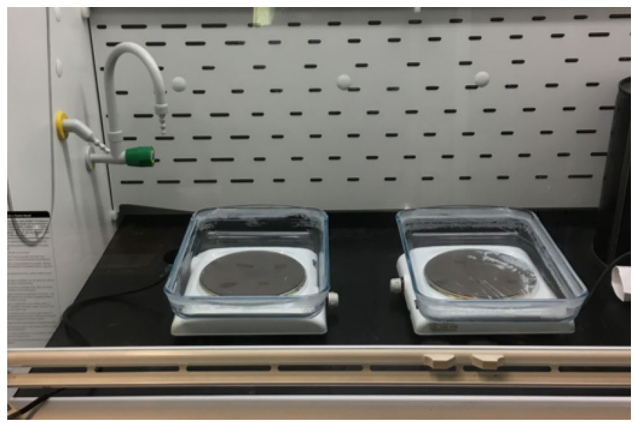

**Fig. 4** Photographs of two Pyrex® trays positioned on top of two electric hobs to achieve the open system reduction

Two electric hobs were turned on and their temperatures were maintained between 55 and 70 °C. With a temperature of 55 °C, the reduction of 100 l of tap water to 1 l took place over 88 h. Tap water was added once per hour to maintain the water level in the trays.

The samples prepared using the open volume reduction technique are listed in Table 2. The uncertainty on the reduction factor $R$ and $\frac{1}{R}$ values results from standard error propagation. Ex: $\frac{\Delta R}{R} = \sqrt{\left(\frac{\Delta \text{Total mass}}{\text{Total mass}}\right)^2 + \left(\frac{\Delta \text{Sample mass}}{\text{Sample mass}}\right)^2}$, where $\Delta R$, $\Delta$Total mass and $\Delta$Sample mass are the uncertainties of the separate quantities $R$, Total Mass and Sample Mass. Sample E was taken after Sample D's reduction had been prepared.

## Results

### Radioassay of water samples

$^{210}$Pb decays via $\beta^-$ decay to $^{210}$Bi, with the most probable $\beta^-$ decay (84%) resulting in an excited state of $^{210}$Bi, emitting an electron with an average decay energy of 4.16 keV. De-excitation of the nucleus results in the emission of a $\gamma$ of 46.5 keV with a 4% yield per decay (Aguilar-Arevalo et al. 2020b). This constitutes the low energy $\gamma$ band, for which HPGe detectors are optimised, therefore justifying the use of these detectors for the radioassay of the water samples.

The Boulby Underground Germanium Suite (BUGS) contains a number of high purity germanium detectors that are discussed in depth in Scovell et al. (2018). For the assay of the samples in this paper, the detector ROSEBERRY

Table 2 London tap water samples

| Sample name | Total mass (g) | Sample mass (g) | Reduction (R) | $\frac{1}{R}$ |
|---|---|---|---|---|
| Tap water sample | n/a | 270 | 1 | 1 |
| Started reduction | 100,685 | n/a | n/a | n/a |
| A | 9926 | 103 | 0.098585 | 10.144 ± 0.001 |
| B | 903 | 102 | 0.008969 | 111.501 ± 0.123 |
| C | 159 | 10 | 0.001579 | 633.239 ± 3.983 |
| D | 28 | 28 | 0.000278 | 3595.893 ± 128.425 |
| E (limescale) | – | 46 | 0.000278 | 3595.893 ± 128.425 |

R in the table denotes the reduction factor calculated as the ratio of the volume after—to the volume before reduction. Total and Sample masses were measured with a set of scales with an uncertainty of 1g. Reduction values have an uncertainty of 0.000009 was used. Roseberry is a Mirion specialty ultra-low background (S-ULB) BE6530 broad energy germanium detector. The planar BEGe detector is constructed to have almost zero dead layer thickness on the front face and, as such, it is optimized for the detection of low-energy gamma-rays. The BE6530 detector has a front face with a surface area of 65 cm² and a thickness of 30 mm. The Roseberry detector is housed in a shield comprising 9cm of lead on the outside, followed by 9 cm copper. The detector cryostat has a j-shaped neck to reduce any line-of-sight background that may come along the aperture through which the detector neck passes through the shielding. This also allows samples to sit on the face of the detector without the need for any additional support.

The sample pots had identical volumes to within manufacturing tolerance, meaning that a single simulation of geometric efficiency could be used. The simulated sample is shown sitting on the head of the Roseberry detector in Fig. 5. A flat spectrum of 0–3 MeV γ-rays are fired from the sample and any which deposit their full energy in the germanium crystal are used to determine the sample efficiency. Figure 6 shows the simulated efficiency with 46.5 keV (corresponding to the $^{210}$Pb γ-ray) highlighted. For the assayed samples, a geometric efficiency of 10.2% was calculated at 46.5 keV.

Samples for the measurements reported here were assayed between August 2019 and January 2020 with runs of lasting between several days for samples that were more radioactive and 2 weeks for samples with little or no $^{210}$Pb above background observed.

## HPGe analysis

Initially a long background run with the detector but no sample pot was performed. Although it was expected that the majority of water samples would be substantially above the detector background this was of particular importance for comparison with the condensate samples from flask B (see

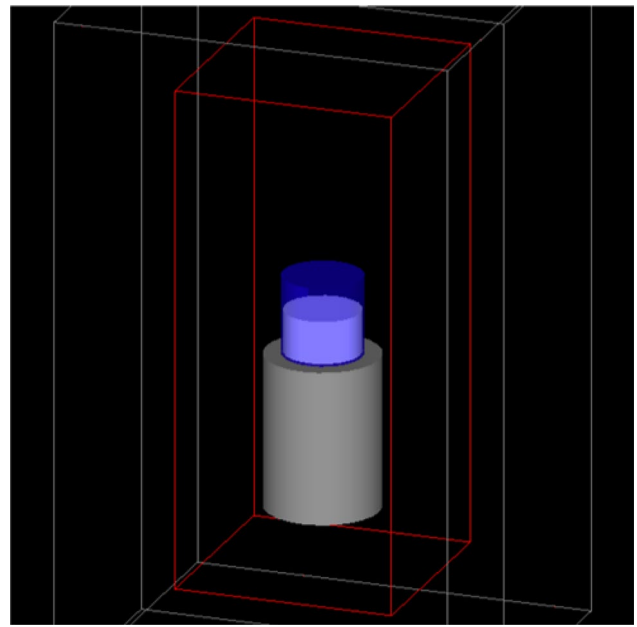

Fig. 5 Simulation showing sample placed on top of the Roseberry detector

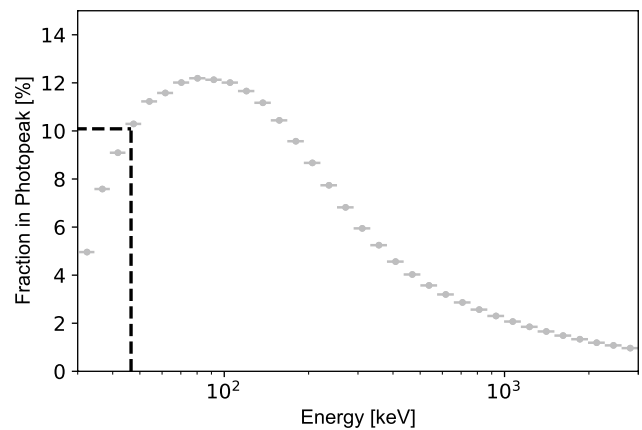

Fig. 6 Spectrum showing simulated efficiency of Roseberry detector. For an energy of 46.5 keV it describes a geometric efficiency of 10.2%, highlighted in the black dotted lines

Table 2) where little, if any, $^{210}$Pb was expected. A background of 20.4 days was used—Fig. 7.

For each sample, a simple and consistent method for determining the number of counts in the $^{210}$Pb peak was used as shown in Fig. 8: the spectral region above and below the peak was fitted using a 0th order polynomial and a simple interpolation between the two fits used to represent the underlying Compton continuum resulting from very small angle scatters of the $^{210}$Pb's 46.5 keV γ-rays (Skinner 1996).

The peak region is defined as shown in Fig. 8 and simple subtraction from the net counts of the Compton background and the measured detector background is performed to give a final number of counts in the peak. Results of this assay programme are shown in Table 3 in comparison with the expected results if the distillation process had 100% efficiency for retaining $^{210}$Pb. Expected Activity $E_A$ is calculated using the mass of the sample and the ratio of activity by the total mass. For sample Red II, for example, $E_A$ is

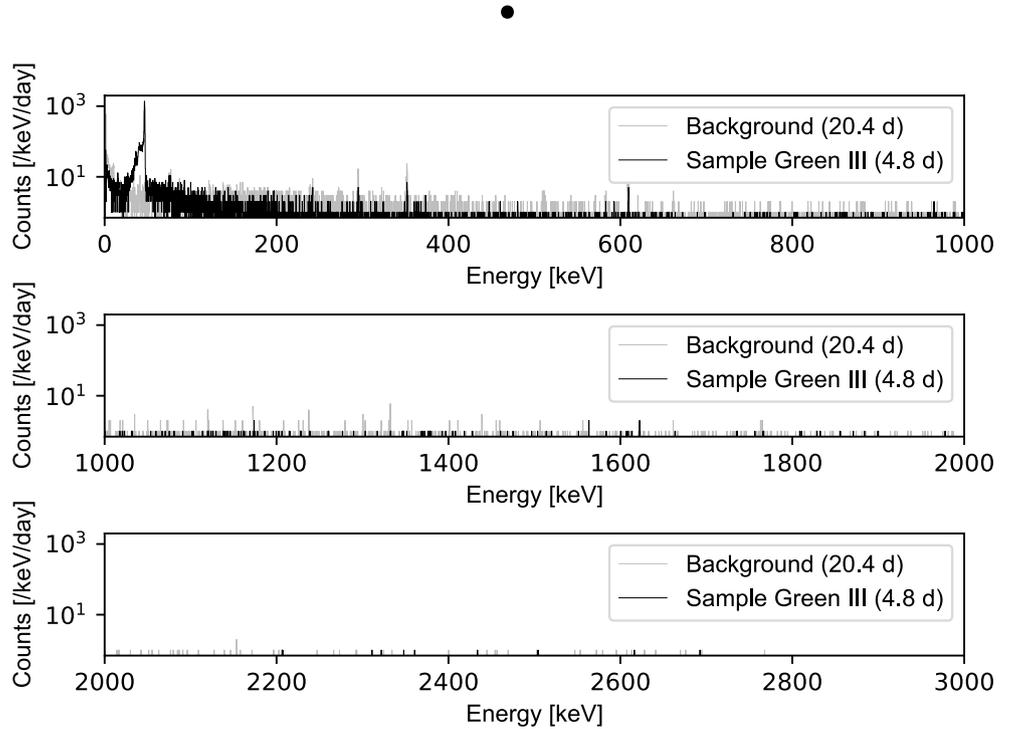

**Fig. 7** Sample Green III—full spectrum for this sample in black, and the background spectrum in grey

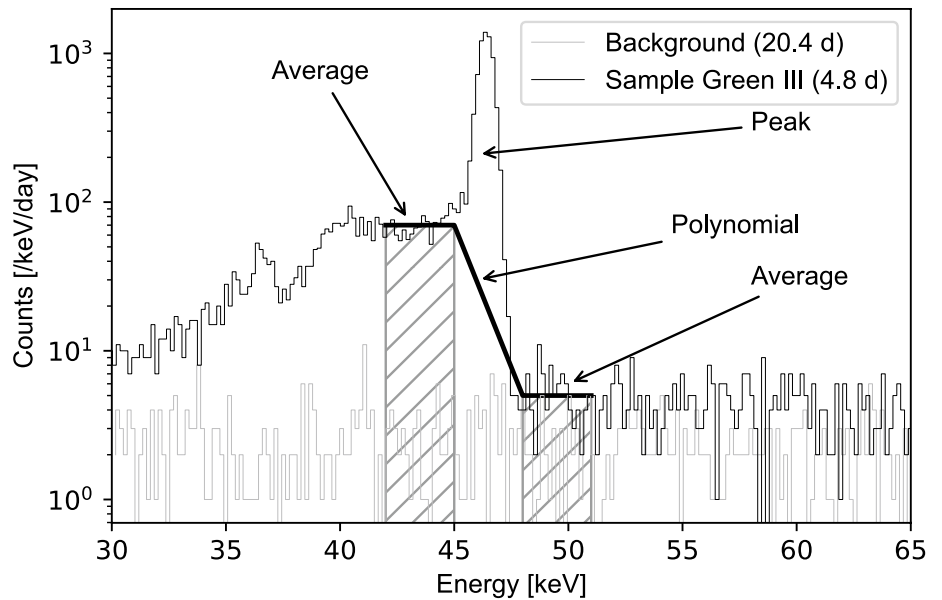

**Fig. 8** Figure 7 zoomed into the region of interest for $^{210}$Pb peak extraction—30–60 keV. The black line represents the 0th order polynomial fits made above and below the peak. The region between 45 and 48 keV represents the peak region

**Table 3** Comparison between measured activity, A, and calculated expected activity, $E_A$, in Bq, assuming >99% efficiency for $^{210}$Pb throughout the volume reduction process

| Sample Name | Measured Activity $A$ | Expected Activity $E_A$ | Ratio $\frac{(A-E_A)}{E_A}$ |
|---|---|---|---|
| Blue I | 0 | 0 | 0 |
| Red II | 5.4 ± 0.4 | 7.9 ± 0.9 | −0.32 ± 0.11 |
| Green III | 4.7 ± 0.3 | 4.6 ± 0.6 | 0.02 ± 0.13 |
| Black IV | 0.006 ± 0.002 | 0 | – |
| Green V | 15.5 ± 1.1 | 13.3 ± 1.8 | 0.16 ± 0.13 |
| Blue VI | 151 ± 8 | 142.1 ± 17.6 | 0.01 ± 0.12 |
| Sample IX | 331 ± 18 | 336 ± 41.6 | 0.01 ± 0.12 |
| Sample X | 0 | 0 | – |
| Total | 507 ± 28 | 503.9 ± 45.2 | 0.01 ± 0.09 |

**Table 4** Cross calibration of sample's activities using BEGe detectors at Boulby and IF-UNAM

| Sample Name | Boulby (stat (Skinner 1996) + sys) | UNAM (stat (Skinner 1996) + sys) |
|---|---|---|
| Blue I | – | 3.4 |
| Red II | 26.9 ± 2.0 | 26 ± 4 |
| Green III | 23.3 ± 1.5 | 17 ± 2 |
| Black IV | 0.029 ± 0.009 | 2.5 ± 0.4 |
| Green V | 77 ± 5 | 71 ± 10 |
| Blue VI | 790 ± 44 | 837 ± 124 |
| Sample IX | 1613 ± 85 | – |
| Sample X | – | – |

All values shown correspond to Specific Activities (Bq/kg)

$200\,g \times \frac{167.9\,Bq}{4236\,g} = 7.9$ Bq, as per the activity and mass values provided in Sect. 2.1. Errors on Activity $A$ combine statistical and systematic effects, the later being estimated by running two simulations using Geant4 and Mirion In Situ Object Counting System (ISOCS) software. The differences in the two efficiency values define the systematic error. Errors on the $E_A$ result from the initial error of the calibration source and the scales used to measure the water sample's mass. Results show that the reduction retains 99 ± 9% of $^{210}$Pb's activity—the total activity $A$ 507 ± 28 Bq compared to an $E_A$ of 503.9 ± 45.2 Bq.

In addition to the assay conducted at Boulby using the ROSEBERRY BEGe detector, the $^{210}$Pb water samples were also characterised at the Institute of Physics of the National Autonomous University of Mexico (IF UNAM) using the BEGe detector described in Aguilar-Arevalo et al. (2020a). The measurements and calibration with both IF UNAM and Boulby Germanium detectors were consistent as described in Aguilar-Arevalo et al. (2020a) and shown in Table 4. For the Boulby measurements, the statistical and systematic effects described for Table 3 also apply here. For the UNAM measurements, a 14% systematic error is included, to account for differences between the real position of the sample and the Monte Carlo simulation. It is calculated by moving the position of the Marinelli Beaker ±4 mm on the axis of the germanium crystal. The water samples contained no isotopes above background, except for $^{210}$Pb. Trace amounts of $^{40}$K were observed for the Blue VI, Green V and Green III reductions, but these were consistent with background.

### Assay of open system water samples

To characterize the efficiency of the open system procedure, samples were assayed between March and April 2019 using the CHALONER BEGe detector at the Boulby Underground Laboratory. CHALONER is a Mirion BE5030 BEGe detector with the same geometry as ROSEBERRY and housed in identical shielding. The intrinsic background of the CHALONER detector is around 10× higher than that of the ROSEBERRY detector (the first is a Very Low Background detector while the later is an Ultra-low Background detector) but, for the samples assayed, this level of background was acceptable. An efficiency simulation was also preformed for CHALONER, identical to was undertaken for ROSEBERRY.

In all samples, $^{40}$K was observed, with an energy peak of 1.46 MeV. The measured gamma energy spectra from this assay are shown in Figs. 9 and 10. As a cross-calibration, the specific activities of $^{40}$K in sample D measured on ROSEBERRY and CHALONER were compared. These activities were 255 ± 6 Bq kg$^{-1}$ (2.83 × 10$^3$ ppb) and 259 ± 2 Bq kg$^{-1}$, respectively. The results of this measurement can be seen in Table 5. Even though Samples A through C show reasonable consistency in activity values, sample D is substantially lower. This is due to the amount of limescale present in this reduction that was removed prior to acquiring the sample (see Table 2). Given that the limescale sample presents an activity equal to 100 Bq/kg, it is reasonable to assume that potassium is lost as limescale is removed and, therefore, the reduction in measured activity of $^{40}$K for sample D is reasonable, as the combined activity of samples D and E is 172 ± 20 Bq/kg. This value is confirmed using the initial measurement of sample D performed on ROSEBERRY. This qualitatively indicates that this method retains at least 90 ± 50% of $^{40}$K, by comparing the average of activity measured in CHALONER from samples A through E, with the initial activity of 167 ± 26 Bq/kg$^{-1}$ (measured in ROSEBERRY).

For the runs on CHALONER, $^{210}$Pb was observed in sample C with a calculated specific activity of 786 ± 369 mBq kg$^{-1}$, which scales to a non-concentrated specific activity of 1.24 ± 0.58 mBq kg$^{-1}$. Although sample D has a concentration factor 5.6× higher than sample C, we see that the Compton continuum of the $^{40}$K peak in this sample was so dominant that the count rate in the expected peak from $^{210}$Pb (at $289 \pm 271$ mBq kg$^{-1}$) is consistent with fluctuations in the background and could easily be lost within it—Fig. 11. This suggests that over-concentration can actually impede the ability to measure $^{210}$Pb in water using a germanium detector.

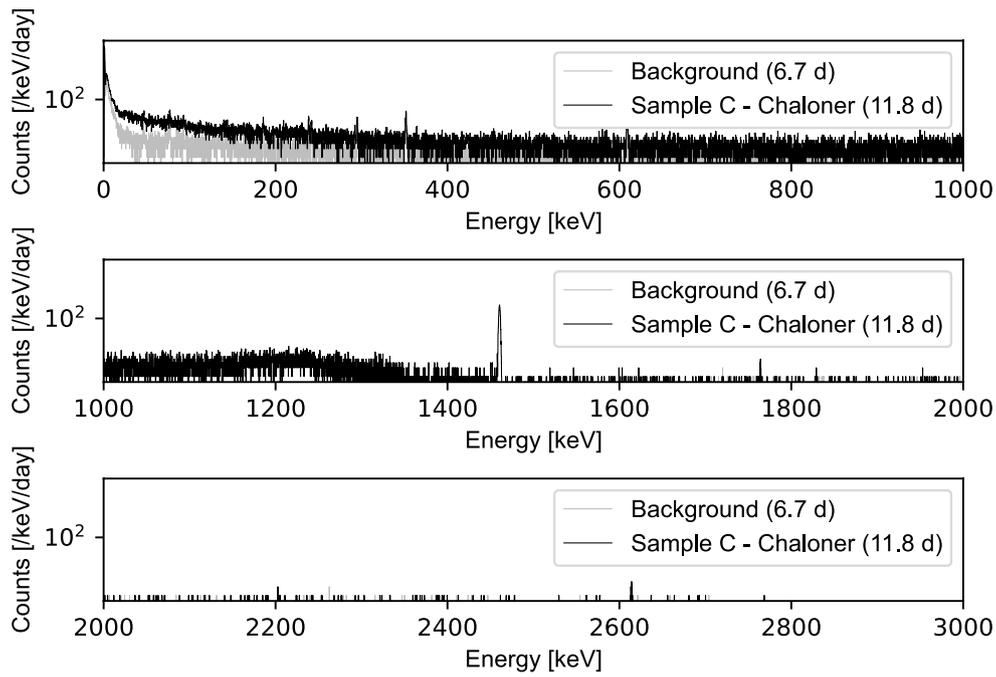

**Fig. 9** Spectra of open water system sample C

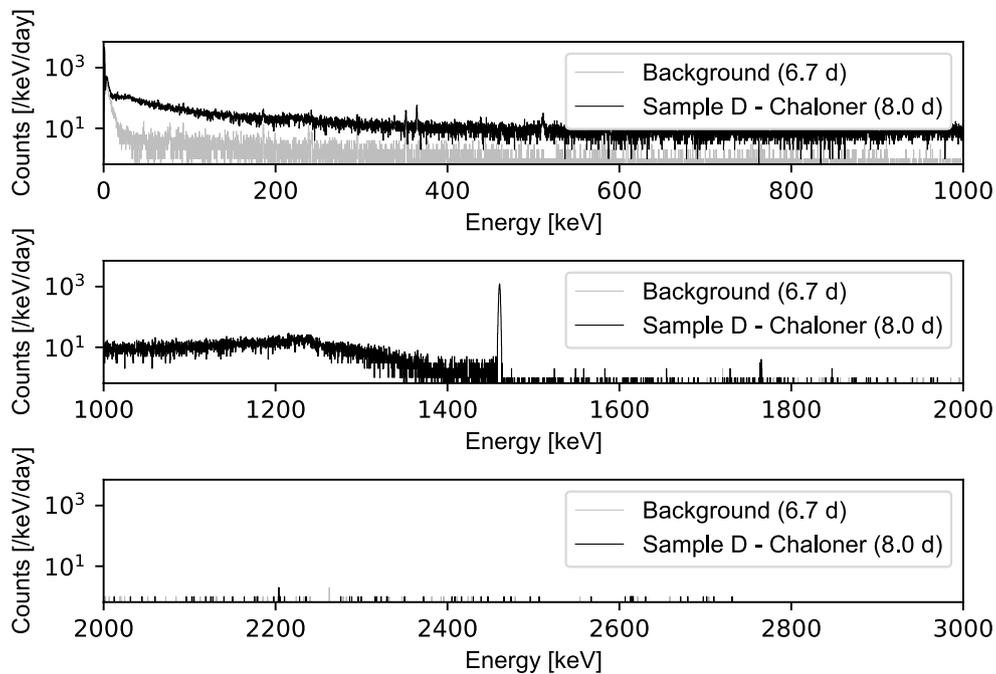

**Fig. 10** Spectra of open system water sample D. It is evident it presents the highest concentration factor out of all assayed samples

From the Roseberry run on sample D, performed on 26 March 2019, a surprising result was observed: peaks in the

**Fig. 11** Spectra of open system water sample D, zoomed to region of $^{210}$Pb peak

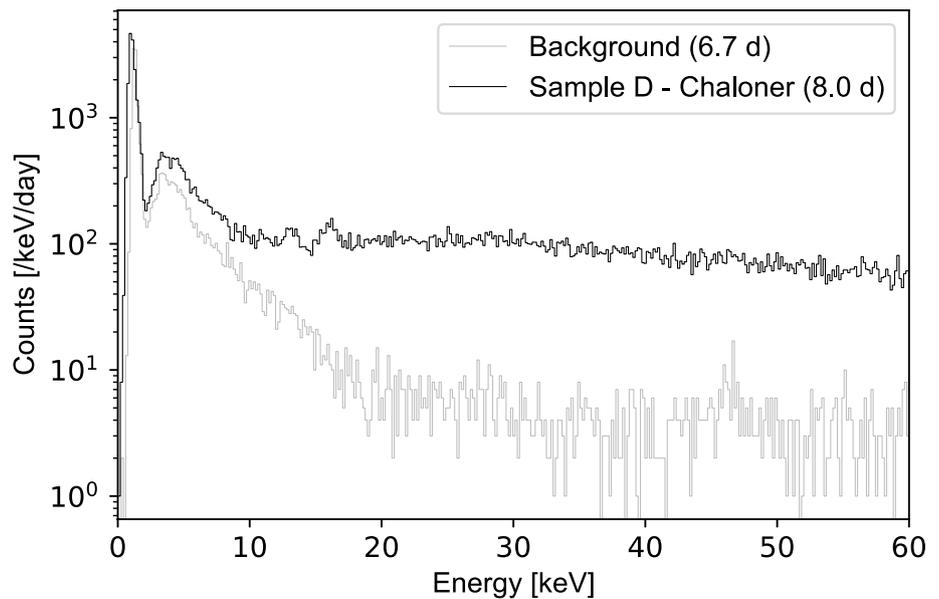

**Table 5** Measured $^{40}$K concentrated specific activities $A$ in all open water samples, in Bq/kg

| Sample name | $A$ (CHALONER) | $A$ (ROSEBERRY) |
|---|---|---|
| Tap water sample | $-8 \pm 106$ | $167 \pm 26$ |
| A | $153 \pm 21$ | – |
| B | $124 \pm 7$ | – |
| C | $150 \pm 13$ | – |
| D | $72 \pm 2$ | $71 \pm 3$ |
| E (limescale) | $100 \pm 20$ | – |

For samples A–D, values are corrected based on the reduction factors described in Table 2. If all $^{40}$K remained in the sample under concentration, all values should be identical. For the tap water sample, the CHALONER run was very short, hence why a high uncertainty is documented energy spectrum consistent with the presence of $^{131}$I were observed with a specific activity of $2.4 \pm 3$ Bq/kg$^{-1}$ ($5.06 \times 10^{-10}$ ppb). During the analysis of the sample D assay using CHALONER (a run starting on the 18 April 2019 and ending on the 1 May 2019) these peaks were re-analysed and were found to have reduced to an average of $0.28 \pm 0.05$ Bq kg$^{-1}$. To verify this result, the decay law, $A = A_0 e^{-\lambda t}$, was used, where $A$ is the number of decays per unit time of a radioactive sample, $A_0$ is the number of decays per unit time at $t = 0$ and $\lambda$ is the decay constant, $= \frac{\ln(2)}{t_{1/2}}$, where $t_{1/2}$ is the half life. Assuming an $A_0$ of $2.4 \pm 0.3$ Bq kg$^{-1}$ and an $^{131}$I $t_{1/2}$ of $8.0252 \pm 0.0006$ d, an estimated reduced activity of $0.20 \pm 0.02$ Bq kg$^{-1}$ was found. This calculation confirms that the measured peaks were consistent with $^{131}$I. This would have been higher at the point the water was extracted from the tap on the 14 March 2019– 5.5 Bq kg$^{-1}$.

In addition to $^{131}$I, $^{177}$Lu was also found in sample D. In the ROSEBERRY run of 26 March 2019, $^{177}$Lu was observed with a specific activity of $3.3 \pm 0.5$ Bq kg$^{-1}$ ($5.84 \times 10^{-10}$ ppb). In the CHALONER run, the activity was shown to have reduced to $0.22 \pm 0.10$ Bq kg$^{-1}$. To verify this result, the decay law was used again. Assuming an $A_0$ of $3.3 \pm 0.5$ Bq kg$^{-1}$ and a $^{177}$Lu $t_{1/2}$ of $6.73 \pm 0.01$d, an estimated reduced activity of $0.17 \pm 0.03$ Bq kg$^{-1}$ was found.

Whilst $^{40}$K occurs naturally and $^{210}$Pb can be found in water in trace amounts, $^{131}$I and $^{177}$Lu are used for medical and industrial purposes. In medicine, $^{131}$I is used in radiotherapy for the treatment of thyrotoxicosis and thyroid cancer (Silberstein et al. 2012; Stokkel et al. 2010). In industry, $^{131}$I is used in the oil industry (Reis 1996). $^{177}$Lu is a component in Lutetium chloride, a radioactive compound used for radio labelling other medicines, such as anti-cancer therapy (European Medicines Agency 2021). In Affinity Water Limited Limited (2021), more than fifty "parameters", including lead, are measured in tap water, but $^{177}$Lu and $^{131}$I do not constitute those parameters. A literature review regarding the presence of these isotopes in London tap water returned only one result (Howe and Lloyd 1986), dated back to 1986, which reports on $^{125}$I and $^{131}$I being measured in environmental samples from the Thames Valley. It is therefore of note that these isotopes can be found in London tap-water.

## Summary and conclusion

This paper reported on two volume reduction techniques, using enclosed and open system procedures. Each method was tested once with a different water sample. We elaborated on the procedures and equipment used, therefore measurements can be repeated by an interested person in similar conditions.

The enclosed system procedure demonstrated that this technique retains $99 \pm 9\%$ of $^{210}$Pb, allowing for an increase of $^{210}$Pb concentration from $1.9 \times 10^{-6}$ ppb to $2.63 \times 10^{-4}$ ppb, or a factor of $1.4 \times 10^{2}$. Given the similarities between $^{210}$Pb and stable lead it can be assumed that stable lead is retained with this efficiency.

The open system procedure showed that all samples contained $^{40}$K with an energy peak of 1.46 MeV. To establish the retention efficiency in the open boiling method, $^{40}$K's concentration was measured across all samples. While some samples showed low variability in activity values, overall the effect of limescale affected these results. It may be of interest to repeat the open water reduction with soft tap water. $^{210}$Pb was observed in sample C (which had a reduction factor of $633 \pm 4$), with a calculated specific activity of $730 \pm 150$ mBq kg$^{-1}$. Using a similar calculation as shown in Aguilar-Arevalo et al. (2020b), this specific activity is equal to $2.55 \pm 0.58 \times 10^{-7}$ ppb. To determine the concentration of stable Pb corresponding to the $^{210}$Pb concentration of sample C, a ratio of 34 ppb of $^{210}$Pb/Pb for rainwater in London was used (Yanga and Appleby 2016). This rainwater ratio can be used for the tap water sample as, according to Thames Water Utilities Limited (2021), rainwater maintains groundwater levels in the region, which consequently supply water to aquifers and rivers, where 70% of drinking water comes from. The ratio allowed an estimation that the sample C had $7.5 \pm 1.7$ ppb of stable lead. Taking into consideration the sample's reduction factor aforementioned, we found that the tap water sample, prior to reduction, contained $0.012 \pm 0.003$ ppb, or $1.2 \pm 0.3 \times 10^{-5}$ mg kg$^{-1}$ of stable lead, a value well bellow the WHO limit. This value is compatible to $< 1$ ppb measured by Affinity Water Limited Limited (2021). When further analysis was done on sample D, $^{131}$I and $^{177}$Lu were also detected, with calculated specific activities of $2.4 \pm 0.3$ Bq kg$^{-1}$ and $3.3 \pm 0.5$ Bq kg$^{-1}$, respectively.

**Acknowledgements**  The Authors wish to thank Ian Murray, Royal Holloway, University of London for his technical support.

**Funding**  This work is supported by the STFC Global Challenges Research Fund (Foundation Awards, Grant ST/R002908/1), by STFC Grant ST/T506382/1, by DGAPA UNAM grants PAPIIT-IT100420 and PAPIIT-IN108020, and by CONACyT Grants CB-240666 and A1-S-8960.

## Declarations

**Conflict of interest**  The authors declare no conflict of interest. The funders had no role in the design of the study; in the collection, analyses, or interpretation of data; in the writing of the manuscript, or in the decision to publish the results.